\documentclass[twoside]{article}
\usepackage{fleqn,espcrc2}

\usepackage{epsfig}

\newcommand{\AmS}{{\protect\the\textfont2
  A\kern-.1667em\lower.5ex\hbox{M}\kern-.125emS}}
\newcommand{\be}{\begin{equation}}
\newcommand{\ee}{\end{equation}}
\newcommand{\bea}{\begin{eqnarray}}
\newcommand{\eea}{\end{eqnarray}}
\newcommand{\MeV}{{\rm MeV}}
\newcommand{\half}{\frac{1}{2}}
\newcommand{\kv}{\vec k}
\newcommand{\mv}{\vec m}
\newcommand{\ev}{\vec e}
\newcommand{\nmunu}{n_{\mu\nu}}
\newcommand{\N} {{\rm N}}
\newcommand{\zmunu}{e^{\frac{2\pi i \nmunu}{\N}}}
\newcommand{\ssu}{\sigma_{\rm SU(2)}}
\newcommand{\sz}{\sigma_{Z_2}}

\title{QCD vacuum structure}                          

\author{M. Garc\'{\i}a P\'erez\address[MCSD]{Theory Division, CERN, 
        CH-1211 Geneva 23, Switzerland}%
}
       
\begin{document}

\begin{abstract}
Several issues related to the structure of the QCD vacuum are reviewed. We
concentrate mostly on results concerning instantons and
center vortices.
\vspace{1pc}
\end{abstract}

\maketitle

\section{Introduction}

The structure of the QCD vacuum has been the subject of many lattice 
investigations over the years. Two phenomena have attracted most  
attention: chiral symmetry breaking and confinement. 
Instantons have been conjectured to play a key role in driving
the former \cite{Shuryak}. 
Here the lattice has entered the game by trying to provide
non-perturbative information on the instanton ensemble.
Activity in this field has already been reviewed in \cite{Tl99,Negele}
and there is not much new to add this year.
This is far from implying that we have reliably
obtained all the information concerning QCD instanton dynamics.
Fundamental issues like the size distribution or density of instantons
are still not really settled,  a point that will be briefly discussed
in section \ref{sec:inst}.

Instantons have also come back on stage due to the beautiful instanton-monopole
link arising at finite temperature \cite{Kvb,Lee}.
This re-establishes some equality between these, a priori, two very 
different objects; instantons are made up of monopoles but 
monopoles can also be viewed as periodic arrays of instantons \cite{vb1}.
 
 Monopoles bring us to the most popular scenario for confinement, 
the dual superconductor picture \cite{Thom}, which
is beautifully at work in SUSY gauge theories \cite{Seibw}.    
't Hooft's proposal of abelian projection \cite{Thoab} has been the subject of extensive
lattice studies. The strongest version of this approach,
based on abelian dominance,  has been criticised in several respects 
\cite{Ddfgo1,Agg,Digia}. 
Different abelian projections do not seem to be equivalent.
Should one of them be preferred, it would imply that a particular U(1)
is selected. The confining flux tube should then be abelian in nature and
fields neutral with respect to it would be unconfined.
A naive particular prediction would be no area law for adjoint Wilson loops. 
Although such
loops are indeed screened at large distances, numerical investigations
indicate the existence of a regime where they exhibit a confining behaviour
with approximate Casimir scaling \cite{Bali}. Moreover, \cite{Ddfgo1} 
one also observes  the more `fundamental' center dominance, where
instead of U(1) the relevant dynamical variables are assumed to be
the ones in the center of the gauge group. Note that center dominance does
not a priori solve the problem of Casimir scaling (adjoint
fields are blind to the center). Although in \cite{Ddfgo1}
this was intended as a criticism, indications towards the possible 
relevance of center vortices to confinement arose from further work 
on the subject \cite{Dfgo2,Kotom}. This has boosted
the revival of a proposal of confinement that for very long remained
asleep \cite{Thov,Mack}. It is based on the fact that center vortices, 
and not U(1) monopoles, are the `confining' configurations. Most of the
activity during the past year has been devoted to this
subject.  I think it would be unfair to review QCD
vacuum structure without acknowledging what has captured most
of the attention. I will thus present my inexpert view on vortices
in section \ref{sec:vortex}.

Results concerning the dual superconductor approach
to confinement will not be reviewed here (for recent reviews see 
\cite{Digia,Chgp}). 
Let me only mention that there is some agreement towards
the fact that abelian dominance is indeed not really the issue
\cite{Digia,Chgp}. Possible resolutions of the puzzles mentioned
before have been proposed.  An important result is that 
one can study condensation of monopoles by constructing a monopole
creation operator \cite{Dglm}. The vev of such an operator is a disorder
parameter for confinement, irrespective of the
abelian projection used to define it. This is considered as a strong
indication in favour of dual superconductivity. Related work concerning 
condensation of magnetic flux will be discussed in section \ref{sec:vortex}.

 The review is structured as follows. I will first concentrate on the 
less speculative phenomena, in particular the calculation of the
topological susceptibility and the $\eta'$ mass. This will be done in 
section \ref{sec:tops}. Section \ref{sec:inst} presents
a few other topics related to instantons, mostly concentrating 
on the instanton-monopole connection at finite temperature
\cite{Kvb,Lee}. In section \ref{sec:vortex} results concerning center
vortices will be presented.

\section{$\chi$ and the $\eta'$ mass}
\label{sec:tops}

A lot of work has been devoted over the past years to computing
the quenched topological susceptibility on the lattice. In the
limit of large number of colours, it is related to the 
$\eta'$ mass through the Witten-Veneziano formula \cite{WV},
\be
\chi_q = \frac{\langle Q^2\rangle}{V} = \frac{f_\pi^2 }{2 N_f} 
\ (m_{\eta}^2 +m_{\eta'}^2 -2 m_K^2) \quad.
\ee
Measuring the fluctuations of topological charge on the lattice
has turned out to be a difficult enterprise (for a detailed description
see \cite{Tl99}). I believe, by now, we can safely say that $\chi_q$ has
been successfully determined both for SU(2) and SU(3). Continuum extrapolations
of the available lattice results give (taking $\sqrt{\sigma}=440\MeV$)
\cite{Tl99}
\bea 
{\rm SU(2)} &&\chi_q = (214 \pm 18 \MeV)^4 \nonumber \\
{\rm SU(3)} &&\chi_q = (200 \pm 18 \MeV)^4 \nonumber 
\eea
in pretty good agreement with the large $\N$ prediction  
$\chi_q \sim (180\MeV)^4$.

The situation is different for the unquenched susceptibility.
In the chirally broken phase 
\be
\chi = \frac{f_\pi^2 m_\pi^2}{2 N_f} + O(m_\pi^4) \propto m_q \quad ,
\label{eq:sust}
\ee
in contrast to the behaviour in the symmetric phase where we
expect $\chi \propto m_q^{N_f}$. The susceptibility is hence an
observable clearly exhibiting the effect of dynamical fermions.
The situation as of Latt'99 did not, however, look that
promising. Available results from CP-PACS and the Pisa group 
\cite{Tl99} failed to see any chiral behaviour and
exhibited an unquenched susceptibility independent of the quark mass.
Results from UKQCD were more encouraging; the expected $m_\pi$ dependence
was indeed observed, although the value of $f_\pi$ extracted from the slope 
of the susceptibility turned out about $20\%$ below the physical value.

\begin{figure}[htb]
\vspace{3.5cm}
\includegraphics{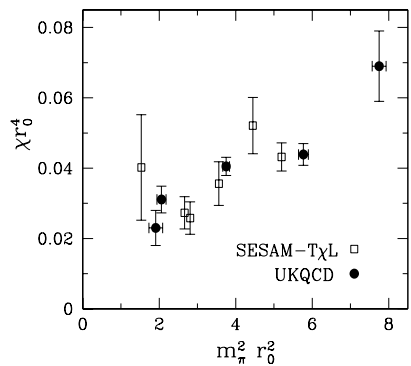}
\caption{Topological susceptibility $\chi r_0^4$ vs $m_\pi^2 r_0^2$.
Comparison between data from \cite{Ukqcd} and \cite{Sesam1}}
\label{fig:sesam}
\end{figure}

New results from UKQCD \cite{Ukqcd}, the Pisa group \cite{Pisa}
and the SESAM-T$\chi$L collaboration \cite{Sesam1} are available.
They use respectively $N_f\!=\!2$ clover improved, staggered and Wilson 
fermions. 
In Fig.~\ref{fig:sesam} a comparison between UKQCD and  SESAM-T$\chi$L's
data for  $\chi r_0^4$ vs $m_\pi^2 r_0^2$ is presented
(with the scale set by Sommer's $r_0\simeq 0.49$ fm). Good
agreement is observed. From a fit to eq. (\ref{eq:sust}) keeping terms
in $m_\pi^4$, UKQCD quotes a value $f_\pi= 105 \pm5^{+18}_{-10}$MeV,
in very good agreement with the expected physical value $f_\pi \simeq 93$MeV.

New results by CP-PACS also indicating the expected 
chiral behaviour have been presented in S. Aoki's plenary talk at
this conference.

\begin{figure}[htb]
\vspace{3.5cm}
\includegraphics{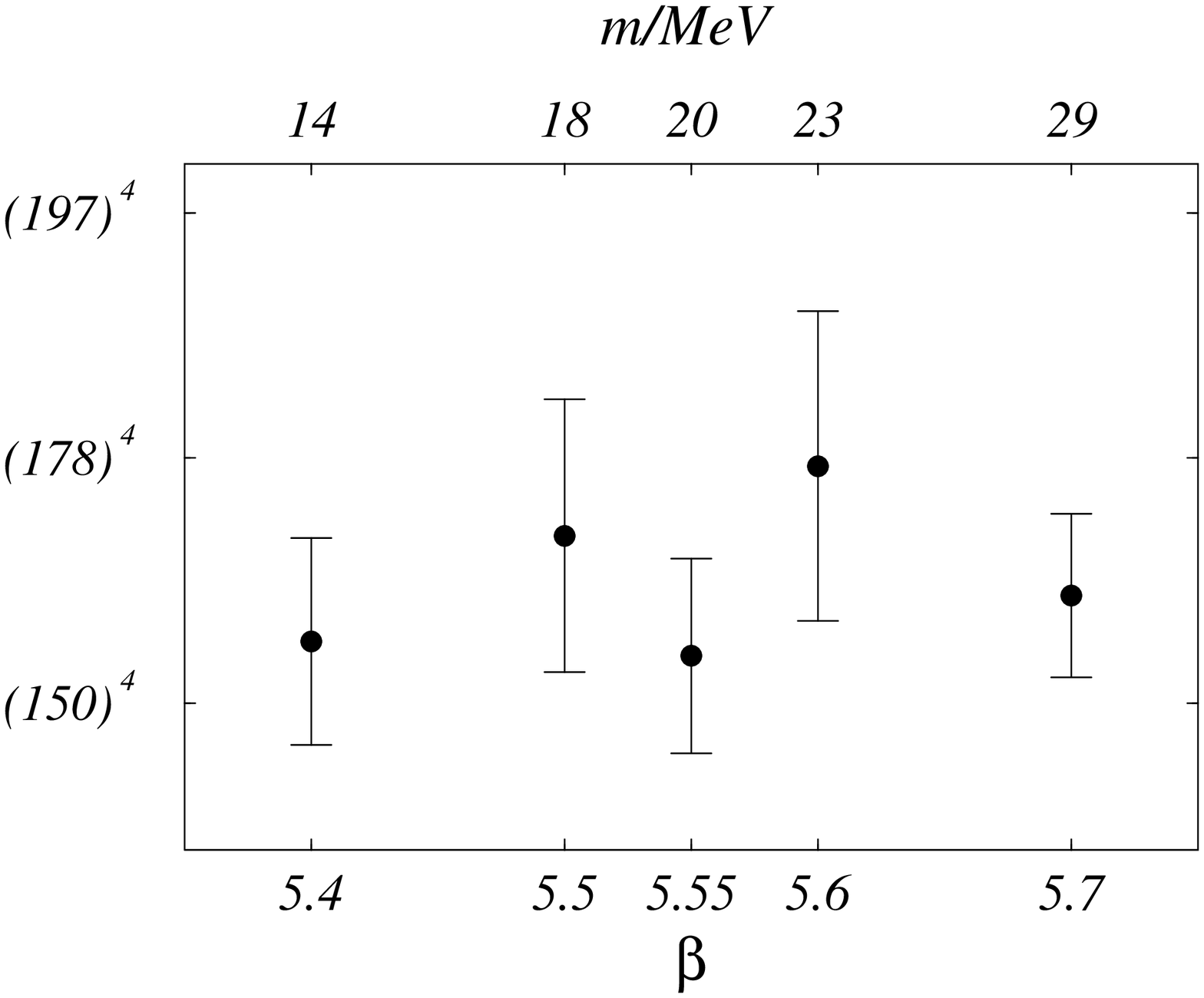}
\caption{$\chi$ vs quark mass from \cite{Pisa}.}
\label{fig:pisa}
\end{figure}

Still, the data from the Pisa group remain puzzling. We present them in
Fig. \ref{fig:pisa}. The errors are rather large, hence it is difficult 
to judge whether the result is really inconsistent with the
expected chiral behaviour. A fit independent of $m_q$ gives 
$\chi = (163\pm 6 \MeV)^4$ with a $chi^2/dof \sim 0.37$, while 
a fit with a linear homogeneous dependence in $m_q$
gives $chi^2/dof \sim 0.94$. The largest 
$\beta$ point, which seems the most problematic, might be affected
by finite size effects which are rather severe for staggered fermions
\cite{Hart}. Taking it out improves the quality of the linear fit
to $chi^2/dof \sim 0.3$. It is in any case clear that improved statistics 
and lighter quark masses are mandatory before deciding if these data
represent a problem.

 To finish this section let me mention some results by
SESAM-T$\chi$L concerning the relation of the $\eta'$ mass 
to topology \cite{Sesam2}. Direct evaluation of the $\eta'$ mass turns 
out to be very tough because it involves OZI suppressed (`disconnected') 
diagrams.  Although improved estimators for these diagrams 
would be required, the connection with topology is already 
clearly exposed in \cite{Sesam2}. 
The ratio between disconnected and connected diagrams is computed in 
sub-ensembles of Monte Carlo configurations characterised by their 
topological charge.  A clear correlation between this ratio and the 
charge is observed. In particular, the disconnected piece vanishes in 
the sub-ensemble with charge $|Q|\le 1.5$, indicating the degeneracy of 
the singlet and non-singlet states in the trivial
topological charge sector.

Results from CP-PACS on the $\eta'$ mass have also been presented in S. Aoki's
plenary talk. 

\section{Instanton constituents}
\label{sec:inst}

 Looking at instantons as composite objects is not a new idea. It is indeed
at work in the non-linear O(3) model in two dimensions - see for instance 
\cite{Poly}- where each instanton seems to be composed 
of a pair of opposite Coulomb charges. It is the melting of instantons into 
a plasma of charges that is argued to induce the mass gap.
In the deconfined phase charges remain bounded in dipoles and long range 
forces are screened.
Similar ideas were put forward for QCD in \cite{Cdg} with fractional topological 
charge configurations, the merons, as fundamental objects. Merons have a
singular action density and turn out analytically intractable. However,
non-singular self-dual objects with fractional topological charge 
exist on a torus with twisted boundary conditions.
They have indeed been advocated as relevant for confinement \cite{Tpa}
and in some sense they can be seen as the fundamental constituents 
of some integer charge instantons.

Recently this constituent nature of instantons has been explicitly
exposed \cite{Kvb,Lee}. The most general finite temperature instanton, 
the so-called caloron, has been constructed. It can be considered as a
periodic array of instantons. For instanton size of the order of their 
separation, each instanton splits into N constituent BPS monopoles.
The splitting takes place when the Polyakov loop at spatial infinity
is non-trivial. Let me, for the sake of simplicity, concentrate 
on the SU(2) case. At spatial infinity the Polyakov loop is constant and 
parametrised by its trace: $2\cos (2\pi w)$. 
The masses of the constituent monopoles are determined by
$w$ and  $\half\!-\!w$. The standard Harrington-Shepard caloron 
with $w=0$ \cite{Hs} is associated to a massive monopole at the center of the 
caloron plus a massless delocalised one at infinity.
It is worth mentioning that the monopoles are located
where the Polyakov loop has two degenerate eigenvalues \cite{vb1}.
This naturally makes contact with abelian projections. 
In a suitable gauge, out of the N monopoles, 
$\N-1$ are BPS static and only one is non-static. 
It is this one which gives topological charge to the caloron. 
Even more, the fermionic zero mode of the $Q=1$ caloron is 
precisely localised on the non-static monopole 
\cite{Zmodes}.

This constituent nature of calorons has already turned out to be
crucial in solving a long standing controversy in SUSY Yang-Mills \cite{Dhkm}.
Calculations of the gluino condensate based on a direct  
semiclassical approximation did not agree with  a
`weak' coupling expression with matter fields added
(and then decoupled) to ensure calculability.
Inclusion of the constituent monopole contributions in the semiclassical
expansion has brought the two expressions to an agreement.

The way to make the constituent monopoles pop-up out of lattice calorons
is by enforcing a non-trivial Polyakov loop at the spatial boundary. 
This is most naturally achieved by introducing twisted boundary conditions
\cite{Caloron1}. Indeed, it can be seen that the SU(2) BPS monopole with mass 
$4\pi^2/\beta$ precisely corresponds with the $Q=1/2$ finite temperature
fractional instanton (allowed to have non-zero magnetic charge due to the
twisted b.c.). 
Non-triviality of the Polyakov loop can also
be achieved by freezing the time links at the spatial boundary.
This has been used in \cite{Impv}  to investigate
the relevant configurations in finite temperature SU(2)
(extracted by cooling). 
The trace of the boundary Polyakov loop is frozen to be zero below $T_c$,
while it is fixed, above $T_c$, to the observed loop average.
In the confined phase, calorons made up of two charge-$1/2$ monopoles 
dominate.  Above $T_c$ such monopoles are still present but
coming in pairs of opposite topological charge.
Dominance of $Q=1/2$ objects for SU(2) at $T=0$ 
has also been observed by imposing magnetic twisted boundary conditions
\cite{Tpa}. Of course the relevant dynamical question, in particular 
below $T_c$, is whether such dominance survives the thermodynamic limit  
irrespective of the boundary conditions used.

The highly non-trivial behaviour of calorons arises due to the 
strong overlap in the periodic array of instantons .
Overlap effects are also important at $T\!=\!0$.
As shown in \cite{Gpkvb} the action density of overlapping instantons 
differs considerably from the simple addition of single instanton profiles. 
Consequences for the extraction of the instanton 
size distribution from the lattice are important. 
In particular, when instantons
are parallel oriented in colour space, large instantons are systematically
missed by instanton finders. This is not an irrelevant issue since a large
instanton component has been argued to give rise to confining behaviour for the
Wilson loop \cite{Diakp}. It is also relevant for the instanton liquid model
\cite{Shuryak} since it indicates a possible failure of the 1-instanton
approximation in ensembles with densities analogous to the ones
obtained in lattice simulations.

A few other works have dealt with this
picture of instanton constituents although from a different point of view.
Ref. \cite{Jzhit} presents a low-energy effective action for QCD that 
incorporates $\theta$ dependence. It can be described in terms of a Coulomb
gas of fractionally charged objects, resembling the constituent monopoles
described above. Also merons have come back on stage in \cite{Sneg}
where regularised lattice merons and their fermionic zero modes
have been obtained.
Finally, ref. \cite{Mdiak} studies whether instantons melt into constituents
in $CP^{(N-1)}$ models with the result that melting
does not take place for $N\ge 3$. 

Let me now mention some other instanton related works.  
In \cite{Ahas,Immp} first evidence for a stronger correlation 
between instantons and anti-instantons in dynamical
configurations has been measured. This has relevance for the instanton 
liquid model which predicts I-A correlations in the presence of 
dynamical quarks \cite{Shuryak}. In \cite{Kovacs} the low-lying mesonic spectrum
has been studied in ensembles of instantons with properties
as obtained from the lattice, much in the spirit of the instanton liquid model.
A first attempt to reproduce the real spectrum failed due to a mixture
between physical states and free lattice modes. Removing these free modes 
by adding a perturbative background results in an excellent agreement.
Finally, the controlled cooling technique developed in \cite{Gpps}
to reduce uncertainties in the analysis of the instanton content of
MC configurations has been extended to SU(3) \cite{Stamw}.

\section{Center vortices}
\label{sec:vortex}

The idea that center vortices might be relevant for confinement in Yang-Mills
theories is also a very old one \cite{Thov,Mack}. It is perhaps in 
3 dimensions 
where it becomes most appealing \cite{Thov}. The 3-D vortex is a topologically
stable soliton of the theory. The vortex creation operator, $\phi$, 
is a local field whose vev signals
the spontaneous symmetry breaking of the $Z_N$ magnetic symmetry, mapping
$\phi \rightarrow e^{\frac{2\pi i n}{\N}} \phi$.
This allows us to postulate an effective low-energy theory in terms of such
a local field.
In the broken phase there are N degenerate vacua, and the Wilson loop
is the creation operator of the domain wall that separates
them.  This wall is stable because the 
vacuum surrounding it is. In this picture the string tension confining 
quarks is related to the tension of the wall.

An extension of these ideas to 4D is not straightforward. In 4D the vortex
creation operator is no longer local (for a recent discussion see
\cite{Kovner}). The 4-D soliton is string-like  
and the natural low-energy effective theory is a theory of strings.
Still, one can, following 't Hooft \cite{Thov}, define quantised electric
and magnetic $Z_N$ fluxes. Consider Yang-Mills theory in a 4-D torus 
of periods $l_\mu$. 
The gauge potential is periodic in $x_\mu$ up to a gauge transformation
$\Omega_\mu(x)$. Univaluedness of $A_\mu$ implies
\be
\Omega_\mu(x+l_\nu)\Omega_\nu(x) = \zmunu \Omega_\nu(x+l_\mu)\Omega_\mu(x)
\ee 
with twist $\nmunu$ (defined mod N) allowed to be non-zero due to 
blindness of $A_\mu$ to $Z_{\N}$.
$m_i\equiv \half \epsilon_{ijk} n_{jk}$ counts the 
magnetic flux in the box along direction $i$, while
$k_i\equiv n_{0i}$ is dual to the electric flux $e_i$. Electric flux 
along a curve $C$ is generated by the Wilson loop, $W(C)$. 
The 't Hooft loop operator, $B(C)$, non-local in the gauge fields,
creates magnetic flux along $C$.  't Hooft argued that in the absence 
of massless particles, the vacuum should be in one of two phases 
parametrised by the vevs of $W$ and $B$. In the Higgs phase they 
show respectively perimeter and area laws. The confined phase is dual 
to it and the roles of $W$ and $B$ are interchanged.

The free energy of a state of given $\ev$ and $\mv$ is
\be
e^{-\beta F(\ev,\mv)} = \sum_{\kv} 
e^{-\frac{2\pi i \kv\cdot\ev}{\N}} Z(\kv,\mv)
\ee
with $ Z(\kv,\mv)$ the partition function 
with twisted boundary conditions specified by $\kv$ and $\mv$. For simplicity
I have assumed that the instanton $\theta$ angle is zero; the $\theta$
dependence gives rise to peculiar phenomena which  will not be 
discussed here.

Euclidean symmetry gives rise to an exact electric-magnetic duality 
between free energies of different fluxes. 
In the presence of a mass gap such duality implies that,  
in the $\beta$, $l_i \rightarrow \infty$ limit,
some of the fluxes have to be heavy. Moreover, 
either all the electric or all the magnetic fluxes are light, 
no mixing between them takes place.
Suppose it is the electric fluxes which are heavy and confined
in thin strings/vortices.
Assuming that at large $\beta$ the free energy factorizes in an electric, 
$F(\ev)$, plus a magnetic, $F(\mv)$, part one derives 
\be
F((m,0,0))=  2\lambda\ (1-\cos(\frac{2\pi m}{\N})) \ l_1\  e^{-\sigma l_2 l_3}
\label{eq:mfree}
\ee
with $\sigma$ the string tension of the electric confining string.
The free energy of magnetic flux decreases exponentially as we let 
the box become large in the plane transverse to the flux. 
Magnetic fluxes spread over the whole volume and condense. 
This behaviour parametrises the confinement phase. 
It is derived solely from duality and the existence 
of heavy electric fluxes.  Indeed, duality does not tell whether it is 
the electric or the magnetic fluxes which condense.
In the Higgs phase the roles of electric and magnetic fluxes are interchanged. 

Implementing twisted boundary conditions on the lattice is rather
easy. For SU(2) a twist $\nmunu=1$ can be 
enforced by flipping the sign of all the plaquettes sitting in, 
for instance, the upper right corner of each $(\mu,\nu)$ plane.
Notice that magnetic flux is only defined modulo 2, as is the number
of twisted plaquettes per plane. 

This year we have seen a revival of calculations of magnetic flux free
energy on the lattice, both at zero \cite{Kovt1,Chel} and at finite
temperature \cite{Hrr,dFdp,Dddgl,Ceac,Hlst}.
Already in \cite{Gjka} it was proposed to use magnetic twist as a probe for
phase structure. In order to compute the free energy
of magnetic flux, one computes the ratio of two partition functions:
$ \exp\{-\beta F(\mv)\} = Z(\mv)/Z(\vec 0)$. 
Results at zero temperature \cite{Kovt1,Chel} support the exponential
behaviour indicated in (\ref{eq:mfree}). It would be nice to check if the
coefficient of the exponential decay of the free energy does indeed
agree with the electric string tension.
At finite temperature there are results, both for 3 \cite{Hlst} 
and 4 dimensions \cite{Hrr,dFdp,Dddgl,Ceac}, corroborating the dual
behaviour of 't Hooft and Wilson loops. There is, for non-zero $T$, 
a difference between introducing the twist in space-time or 
space-space planes. As shown in \cite{dFdp} space-space 
't Hooft loops show area law, corresponding 
to  deconfinement of space-time Wilson loops. However, space-time 't 
Hooft loops are screened in agreement with the observation that the spatial 
string tension survives above $T_c$. Similar results are obtained in 
\cite{Hrr,Hlst}. Previous analytic calculations of the expectation value of 
the 't Hooft loop at high temperature \cite{Kakovs} also support this
picture. 

It is worth mentioning the comparison  in \cite{Dddgl,Ceac}
between disorder operators signalling monopole and magnetic flux
condensation. Both of them behave in a very similar way giving a
critical temperature compatible with the standard determinations.

Another issue much more difficult to settle, is whether indeed the
disordering of Wilson loops is driven by the presence of thick magnetic
vortices in the vacuum as advocated in \cite{Mack}. 

Smooth vortex configurations do exist and have been obtained on 
the lattice from cooling \cite{Gam}. 
For this the use of twisted
boundary conditions is again essential. Here one remark
is important. With twisted b.c. such that $\kv\cdot \mv \ne 0$ 
(mod N) the topological charge is fractional and quantised 
in units of $1/\N$.  We have already discussed fractional charge 
objects in connection to monopoles. 
Vortices found in \cite{Gam} also carry fractional charge. 

Another nice connection between vortices and topology has been put forward in
\cite{Engel}. Based on a model that describes vortices as random
surfaces \cite{Engelr}, topology is incorporated by providing orientation
to the surface. Non-trivial topological charge comes from 
non-globally-orientable surfaces describable as patches of 
equal orientation separated by monopole lines. A prediction for the
zero temperature susceptibility of $\chi_q (T\!=\!0) \!=\! 
( 190\pm 15 \MeV)^4$ is derived, in amazingly good agreement with 
lattice results - see sec. \ref{sec:tops}.

How to locate thick center vortices on lattice configurations
has been the subject of a big debate this year. In \cite{Dfgo2} an approach 
very similar to 't Hooft's abelian projection was taken (other
alternative approaches will not be discussed, for a recent review see
\cite{Ktrev}). Center vortices are located by fixing the  
so-called maximal center gauge (MCG), obtained by maximising
the average of $ | {\rm Tr} (U_\mu(x)) |^2$.
Let us concentrate on the SU(2) case. Center projection consists of 
replacing gauge fixed links by the closest $Z_2$ element.
Center-projected (P-)vortices correspond to coclosed
sets of plaquettes taking value (-1).  It is claimed that
the string tension from center-projected links ($\sz$) 
agrees with the full string tension ($\ssu$),
a phenomena dubbed as center dominance. 
The relevance of this is, however, obscured by the fact that center 
dominance appears to be obtained even without gauge fixing \cite{Fgoo2}. 
The physicality of P-vortices has to be judged on a different basis. 
Tests in \cite{Dfgo2} correspond to the behaviour
of Wilson loops pierced by even/odd number of P-vortices, and to the
scaling of the P-vortex density. The behaviour of P-vortices across
the deconfinement phase transition has also been studied \cite{Lter}.

The news this year is that maximal center gauge turns out to be 
severely affected  by lattice Gribov copies. A first indication in 
this direction was provided in \cite{Kovt2}. If, prior to fixing MCG,
the configurations are driven into a smooth gauge like the
Landau gauge, center dominance is lost and the
density of P-vortices dramatically reduced. Worrisome 
is that Landau preconditioning usually gives a higher
local maximum than the one from
direct MCG. Further evidence comes from
\cite{Bkpv,Bkp} where several random copies of 
the same configuration are made, taking from them the one that gives the 
higher local maximum after MCG. The number of copies is
extrapolated to infinity, with the result again that a very significant
part of $\ssu$ is lost in center projection, 
even in the continuum limit. The debate originated 
about this issue (see \cite{Bkpv,Bfgo}) seems settled in
\cite{Bkp} with a very careful study of the dependence on the
number of gauge copies and the results stated above.

 One can do better by performing a gauge fixing free of lattice 
Gribov ambiguities. This is the case of the Laplacian gauge,  
first introduced for abelian projection \cite{Vinkw} and further 
extended to perform center gauge fixing \cite{dFp1}
(see also \cite{Tok}).  
The idea is to diagonalise the adjoint Laplacian
and use its two lowest eigenvectors to fix the gauge. Here instantons,
monopoles and vortices arise respectively as point, 1-
or 2-dimensional singularities of the gauge fixing \cite{dFp2}. 
Indeed, in the Laplacian gauge, center dominance is recovered, although only 
in the continuum limit. But we have by now repeatedly said that center 
dominance alone is not good enough. Further investigation on the vortex content 
of Laplacian gauge fixed configurations is still necessary.

\section{Closing remarks}

There is still a lot of work to do to unveil the mysteries 
of confinement.  Perhaps the relative failure of the approaches taken so far 
is related to our insistence on describing confinement in 
`semiclassical' terms. We tend to bear in mind that some 
underlying `classical' fields (be it `fat' monopoles, 
vortices or instantons) drive the phenomena. But attempts to identify
them in non-perturbative ensembles have systematically led
to problems. There might be some truth in it but the key ingredient
seems to be still missing.

\section*{Acknowledgements}
I thank the organisers for a very enjoyable conference.
I am indebted to Jos\'e Luis F. Barb\'on, Tony Gonz\'alez-Arroyo and 
Pierre van Baal for invaluable discussions over the 
years. I would also like to thank Philippe de Forcrand, Adriano Di Giacomo, 
Jeff Greensite, Tamas Kovacs, Alvaro Montero, Carlos Pena and Nucu Stamatescu.

\end{document}